\documentstyle[aps,twocolumn]{revtex}
\draft
\begin{document}
\author{V.V.Flambaum,$^{1,2}$
F.M.Izrailev$^{1,2}$
}

\small
\address{$^1$ School of Physics, University of New South Wales,
Sydney 2052, Australia}

\address{$^2$ Budker Institute of Nuclear Physics, 630090 Novosibirsk,
Russia}

\title{Distribution of occupation numbers in finite Fermi-systems and\\
role of interaction in chaos and thermalization}

\date{\today}
\maketitle

\begin{abstract}
New method is developed for calculation of single-particle
occupation numbers in finite Fermi systems of interacting particles. It is
more accurate than the canonical distribution method and gives the
Fermi-Dirac distribution in the limit of large number of particles. It is
shown that statistical effects of the interaction are absorbed by an increase
of the
effective temperature. Criteria for quantum chaos and statistical equilibrium
are considered. All results are confirmed by numerical experiments in the
two-body random interaction model.
\end{abstract}

\narrowtext

Studying numerous problems related to many-body compound states one
needs to know whether the laws of statistical physics can be used in the
description of a particular quantum system with finite number of interacting
particles (nucleus, atom, etc). The aim of the present work is to develop
an accurate method for calculation of occupation numbers (and other statistical
average values) in finite isolated systems and  compare the results
with the Fermi-Dirac (FD) distribution. The latter is a priori valid in
infinite systems of non-interacting particles. We show that the accuracy of
 the FD distribution in finite
systems of interacting particles can be improved by introduction of the
effective temperature which absorbs statistical effects of the interaction.  

For finite systems of interacting Fermi-particles the occupation
numbers $n_{s}$ of single-particle orbitals can be found if we know the
expansion of exact eigenstates $\left| i\right\rangle $ in terms of Slater
determinants (``shell model states'') 
\begin{eqnarray}\label{slat}
n_s=\left\langle i\right| \hat n_s\left| i\right\rangle =\sum\limits_k\left|
C_k^{(i)}\right| ^2\left\langle k\right| \hat n_s\left| k\right\rangle \\
\label{slat1}
\left| i\right\rangle =\sum\limits_kC_k^{(i)}\left|
k\right\rangle \,;\,\,\,\,\,\,\,\,\left| k\right\rangle
=a_1^{\dagger}\dots a_n^{\dagger}\left| 0\right\rangle 
\end{eqnarray}
Here, $\hat n_s=a_s^{\dagger}a_s$ is the occupation number operator. 

 If the residual interaction between the particles is strong enough,
the expansion (\ref{slat1}) can be treated as a ``chaotic'' superposition of
the basis states $\left| k\right\rangle $ . In this case mean square
 values of the
expansion coefficients are smooth functions of the energy difference between
the energy $E_k$ of the basis state $\left| k\right\rangle $ and the energy $%
E^{(i)}$ of the exact state $\left| i\right\rangle $, $\overline{\left|
C_k^{(i)}\right| ^2} \equiv F_k^{(i)}=F^{(i)}(E_k-E^{(i)})$. 

 In numerical studies of the Ce atom \cite{ce}, the $s-d$ nuclear
shell model \cite{ZELE} and random two-body interaction model \cite{FGI96},%
\cite{FIC96} it was demonstrated that typical shape of exact eigenstates
(``spreading function'') $ F_k^{(i)}\,$practically does not depend on
 a particular
system and has a universal form characterized by the spreading width $\Gamma 
$ . For example, for Ce atom \cite{ce} the squared Breit-Wigner shape of $%
 F_k^{(i)} $ has been found in a good agreement with numerical data, $%
 F_k^{(i)}\propto [ ( E_k-E^{(i)}-\Delta _1^{(i)}) ^2+ 
\Gamma ^2/4 ] ^{-2}$ , where $\Delta _1^{(i)}\ll \Gamma $ is a
small shift (see below) which in the zero approximation can be neglected.
The value of $\Gamma =2\left[ \overline{\left( \Delta
E\right) ^2}\right] ^{1/2}$ can be found using the following relations for
the spreading width of basis components, 
\begin{equation}
\label{delE}\overline{\left( \Delta E\right)_k^2}\equiv \sum\limits_i\left|
C_k^{(i)}\right| ^2\left( E_k-E^{(i)}\right) ^2=\sum\limits_{p\neq
k}H_{kp}^2 
\end{equation}
with $H_{kp}$ standing for non-diagonal Hamiltonian matrix elements defined
by residual interaction $V$. 
For example, in the model of $n$ particles distributed over $m$ orbitals
 we have \cite{FGI96} $ (\Gamma/2)^2=\overline{\left( \Delta E\right) ^2%
}=V_0^2 n(n-1)(m-n)(3+m-n)/4$. 
Here $V_0^2=\,\overline{\left| V_{st\rightarrow pq}\right| ^2}$ is the mean
square value of two-body residual interaction matrix elements.
 
Thus, we can
rewrite Eq. (\ref{slat}) in the form of the `` $F-$distribution'' which gives
the actual distribution of occupation numbers in finite Fermi systems, 
\begin{equation}
\label{nalpha}n_s(E)=\frac{\sum\limits_kn_s^{(k)}\,F\,(E_k-E)}{%
\sum\limits_k\,F\,(E_k-E)} 
\end{equation}
$$
E_k=H_{kk}=\sum\limits_sn_s^{(k)}\,\epsilon
_s+\sum\limits_{s>p}u_{sp}n_s^{(k)}n_p^{(k)} 
$$
where $n_s^{(k)}\equiv \left\langle k\left| \hat n_s\right| k\right\rangle $
equal $0$ or $1,$ $\epsilon _s$ are the energies of single--particle
orbitals, and $u_{sp}=V_{sp\rightarrow sp}$ is the diagonal matrix element
of residual interaction. In practice, the second term in the definition of $%
E_k$ can be substantially reduced by an appropriate choice of the mean
field. The denominator in (\ref{nalpha}) stands for the normalization of the 
$F-$distribution. This ``microcanonical'' distribution is convinient for
numerical calculations. 

 It is instructive to compare the $F-$distribution (\ref{nalpha})
with occupation numbers obtained by making use of the standard canonical
distribution, 
\begin{equation}
\label{gibbs}n_s(T)=\frac{\sum\limits_in_s^{(i)}\,\exp (-E^{(i)}/T)}{%
\sum\limits_i\exp (-E^{(i)}/T)} 
\end{equation}
were $T$ is the temperature. The difference between Eqs. (\ref{nalpha}) and
(\ref{gibbs}) is that the summation in Eq. (\ref{nalpha}) is performed over
simple basis states while in the canonical distribution the summation is
carried out over exact eigenstates. Another difference is that in Eq.
(\ref{nalpha}) the occupation numbers are calculated for a specific energy $E$
of system unlike specific temperature $T$ in Eq.(\ref{gibbs}). However, Eqs.
(\ref{gibbs}) and (\ref{nalpha}) can be compared with each other by setting
 $E=\left\langle E\right\rangle _T$. 

 It is important to demonstrate that the $F-$distribution (\ref
{nalpha}) tends to the standard FD distribution in the limit of
large number of particles. By performing explicitly the summation over $%
n_s=0 $ and $n_s=1$, the expression (\ref{nalpha}) can be written in the
form 
\begin{equation}
\label{nsZ}n_s(E)=\frac{0+Z_s(n-1,E-\tilde \epsilon _s)}{Z_s(n-1,E-\tilde
\epsilon _s)+Z_s(n,E)} 
\end{equation}
Here, the partition function is introduced, $Z_s(n,E)=\sum\nolimits_k^{%
\prime }F\,(E_k-E)$ , where the summation is taken over all states of $n$
particles with the orbital $s$ excluded. Correspondingly, the sum in $%
Z_s(n-1,E-\tilde \epsilon _s)$ is taken over the states of $n-1$ particles
with the orbital $s$ excluded; this sum appears from the terms for which the
orbital $s$ is filled $(n_s=1)$. For such states one can write $%
E_k(n)=\tilde \epsilon _s+E_k(n-1)$ where $E_k(n-1)$ is the energy of the
basis state with $n-1$ particles and $\tilde \epsilon _s=\epsilon
_s+\sum\limits_{p\neq s}u_{sp}n_p^{(k)}$. Note that to add the
energy $\tilde\epsilon _s$ to  $E_k(n-1)$  is the same as to
subtract it from $E$ because $F=F(E_k+\tilde \epsilon _s-E)$.

 By taking $\tilde \epsilon _s$ independent on $k$ we assume the
averaging over the basis states near the energy $E$ is possible. Number of
terms $N$ in the partition function $Z_s\,$ is exponentially large, $N=\frac{%
m!}{(m-n)!n!}$ , therefore, one should consider $\ln \,Z_s$ which has slow
dependence on $n$ . In the case of large number of particles one can
get $\ln \,Z_s(n-\Delta n,E-\tilde \epsilon _s)=\ln \,Z_s(n,E)-\alpha
_s\,\Delta n\,-\beta _s\tilde \epsilon _s$ where $\alpha _s=\frac{\partial
\ln \,Z_s}{\partial n};\,\,\,\,\,\,\beta _s=\frac{\partial \ln \,Z_s}{%
\partial E};\,\,\,\,\,\,\Delta n=1$ . This leads to the ``FD'' type
of the distribution, 
\begin{equation}
\label{FDtype}n_s=\left( 1+\exp (\alpha _s+\beta _s\tilde \epsilon
_s)\right) ^{-1} 
\end{equation}
If the number of substantially occupied orbitals in the definition of $Z_s\,$
is large, the parameters $\alpha _s$ and $\beta _s$ are not sensitive as to
which particular orbital $s$ is excluded from the sum and one can assume $%
\alpha _s=\alpha \equiv -\mu /T,\,\,\beta _s=\beta \equiv 1/T$ . The
chemical potential $\mu $ and temperature $T$ can be found from the
conditions 
\begin{equation}
\label{eqs}\sum\limits_sn_s=n;\,\,\,\,\,\sum\limits_s\epsilon
_sn_s+\sum\limits_{s>p}u_{sp}n_sn_p=E 
\end{equation}

 In the case of many non-interacting particles (ideal gas) similar
procedure transforms the canonical distribution (\ref{gibbs}) to the
FD distribution (see e.g. \cite{book}). It is easy to check that
the canonical distribution coincides with the FD distribution with
a high accuracy even for very small number of particles provided the number
of effectively occupied orbitals is large (when $T\gg \epsilon $ or $\mu
\gg \epsilon $) . However, for the fixed total energy $E$, the
temperature $T$ in these two distributions is different. This difference can
be explained by the dependence of the parameter $\alpha _s$ on $\epsilon _s$
(see Eq. (\ref{FDtype})). Indeed, using expansion $\alpha _s=\alpha
(\epsilon _F)+\alpha ^{\prime }(\epsilon _s-\epsilon _F)$ one can obtain the
relation between the FD ($\beta _{FD})$ and canonical ($\beta )$
inverse temperatures: $\beta _{FD}=\beta +\alpha ^{\prime }\epsilon _F$. 

 Now, we can study the accuracy of the distributions (\ref{nalpha},%
\ref{gibbs},\ref{FDtype}) in the description of realistic quantum systems.
As is known, the FD distribution is valid if the gas of particles
can be considered as ideal, therefore, when the residual interaction is
small enough, $\Gamma $ $\ll \mu \sim nd_0$ or $\Gamma $$\ll T$ (here $d_0$
is the mean energy spacing for single-particle states). It is also assumed
that the number of particles is large, $n\gg 1$. However, the equilibrium
distribution for occupation numbers arises for much weaker condition,
namely, when the number of principal components $N_{pc}$ in exact
eigenstates (see (\ref{slat1})) is large, $N_{pc}\sim \Gamma /D\gg 1$ . In
this case the fluctuations of the occupation numbers $\sim $ $N_{pc}^{-1/2}$
. Since the energy interval $D$ for many-body states is exponentially small,
it is enough to have relatively weak residual interaction $V_0\gg D\sim
d_0\,exp(-n)$ in order to get the equilibrium distribution. 

 There are four regions of parameters depending on the strength of
interaction and number of particles: 

 (I) ``regular'' states, $N_{pc}\approx 1$ for $V_0<D$  \ ; 

 (II) ``initial chaotization'' which is characterized by a relatively
large number of ``random'' principal components, say, $N_{pc}\sim \Gamma
/D\geq 10$, however, the fluctuations are still large, $N_{pc}^{-1/2}\geq
0.1 $ ; 

 (III) equilibrium $F-$distribution (\ref{nalpha}) which is
characterized by small fluctuations, $N_{pc}^{-1/2}\ll 1$ , or, $N_{pc}\sim
\Gamma /D\geq 100;$ in this case components of eigenstates can be treated as
random variables with the variance $F$ and  the Eq.(\ref{nalpha})
gives actual distribution of the occupation numbers in quantum systems with
interacting particles; 

 (IV) canonical distribution (\ref{gibbs}) which arises in the case
of equilibrium plus large number of particles. If, in addition, the
condition $\Gamma \ll nd_0$ is fulfilled, the standard FD
distribution is valid. 

 In practice, the condition (IV) of ``thermalization'' is not easy to
satisfy in realistic systems like atoms or nuclei since $n$ in this estimate
is, in fact, the number of ``active'' particles (number of particles in the
valence shell) rather than the total number of particles. Thus, the
equilibrium $F-$distribution (\ref{nalpha}) which does not require the
thermalization condition (IV) is more accurate. 

 To test the above statements we have performed a detailed numerical
study of the model of two-body random interaction. This model (see details
in \cite{FGI96},\cite{FIC96}) is described by few parameters: number $n$ of
particles, number $m$ of orbitals and ratio $V_0/d_0$ of the two-body
interaction strength to the spacing between single-particle levels. For very
small interaction the eigenstates are ``regular'' and the occupation number
distribution $n_s$ is a strongly fluctuating function even after averaging
over a number of close eigenstates, see Fig.1. With an increase of the
interaction keeping the number of particles small, we obtain equilibrium $%
F- $ distribution which is different from the FD
distribution (see below). 

 If, instead, we increase the number of particles keeping the
interaction small, $V_0\ll d_0$ , the distribution (\ref{nalpha}) tends to
the FD one (Fig. 2a). Finally, when the strength of the interaction
is beyond the ideal gas approximation, the equilibrium distribution strongly
deviates from the standard FD distribution (Fig. 2b). In fact, the
latter result happens for a relatively small interaction $V_0\sim 0.1d_0$
which, however, results in the large value of spreading width, $\Gamma >d_0$
since $\Gamma $ increases with the number of particles very fast. 

 However, the accuracy of the FD distribution can be
improved by renormalization of the temperature. The point is that
statistical effects of the interaction can be, at least in part, described
by the increase of effective temperature. For the first time, it was pointed
out in Ref. \cite{FIC96} where we demonstrated that the spreading widths $%
\gamma $ of single-particle orbitals are imitated by the increase of
temperature, namely, $n_s(\epsilon _s,\gamma ,T)\approx n_s(\epsilon
_s,\gamma =0,T+\Delta T)\equiv n_s(\epsilon _s,\tilde T)$; for $\gamma \ll
\mu \,$ one can obtain $\Delta T\approx \gamma ^2/16T$. Below we present
stronger arguments. Since we are mainly interested in the eigenstates which
are in the lower half of the spectrum, the mean field energy (\ref{eqs}) is
substantially higher than the actual energy of eigenstates $E^{(i)}$. This
results from level repulsion which forces lower energies to shift down. The
mean field energy does not ``know'' about the level repulsion (the shift
appears in the second order of perturbation theory), therefore, the
unperturbed energy of the level with the same number would be a better
estimate for it ($E=H_{ii}$ instead of $E=E^{(i)}$). Thus, we should
substitute into Eq. (\ref{eqs}) the energy $E$ which is higher than the
energy of eigenstate: $E=E^{(i)}+\Delta _1^{(i)}\simeq H_{ii}$. 

 There is one more effect which leads to an additional shift $\Delta
_2^{(i)}$ . Since the density of states rapidly increases with the energy,
the number of admixed higher basis states is larger than that of the
lower states (an extreme example is the ground state which contains the
admixture of the higher basis components only). It is easy to estimate the
corresponding increase of the mean field energy due to this effect if the
spreading function $F(E_k-E^{(i)})$ is symmetric (to separate the effects
we set here $\Delta _1^{(i)}=0$): 
\begin{equation}
\label{shift}\Delta _2^{(i)}=\int F_k^{(i)}(E_k-E^{(i)})\rho
_0(E_k)dE_k\simeq \frac{d(\ln \rho _0)}{dE}\overline{\left( \Delta E\right) ^2}
\end{equation}
Here $\rho _0$ is the unperturbed
level density, and $\overline{\left( \Delta E\right) ^2}$ is given by Eq.(%
\ref{delE}). According to \cite{FW70}, the shape of density of states for $%
m\gg n\gg 1$ is close to the Gaussian both for non-interacting and
interacting particles,
 with the center of the spectrum $E_c$ and variances 
$(\sigma _E)^2$ and $(\tilde \sigma _E)^2= (\sigma _E)^2+
\overline{(\Delta E)^2}$ correspondingly. Therefore, one can obtain simple
 estimates of the shifts: 
\begin{equation}
\label{shifts}\Delta _1^{(i)}\simeq \left( E_c-E^{(i)}\right) \frac{
\overline{\ (\Delta E)^2}}{2(\sigma _E)^2}\,;\,\,\,\,\,\,\,\,\,\,\,\Delta
_2^{(i)}\simeq 2\Delta _1^{(i)} 
\end{equation}
  Taking into account these energy shifts and Eq.(\ref{eqs}) one can
estimate the increase of the effective temperature of the Fermi gas due to
interaction. An accurate calculation of this effect requires more detailed
knowledge of the spreading function $F$ including it's weak ``non-resonant''
energy dependence on $E_k$ and $E^{(i)}$. However, the occupation numbers
determined by the $F-$distribution (\ref{nalpha}) are not sensitive to a
particular choice of the spreading function provided the condition $\sqrt{%
N_{pc}}\gg 1$ is fulfilled. To check this statement we have considered the
form of the spreading function which takes into account important features
of actual distribution $F(E)$ . If the interaction is small, in the region
not very far from the maximum, $F_{\max }\sim D/\Gamma \propto \rho ^{-1}$,
the spreading function $f(E)\,\,$ can be described by the Breit-Wigner form
with the spreading width $\Gamma _{BW}=2\pi \overline{V^2}/D$ \cite{BM}.
Also, there is a shift of the maximum due to the repulsion between
neighboring levels. These arguments allow us to suggest the improved
expression for the spreading function valid for $\Gamma _{BW}<\Gamma $:

\begin{equation}
\label{newBW} F_k^{(i)} \propto \frac{\left( \rho _0(E_k)\rho
(E^{(i)})\right) ^{-1/2}}{\left[ \left( E_k-E\right) ^2+\frac{\Gamma _1^2}%
4\right] \left[ \left( E_k-E\right) ^2+\frac{\Gamma _2^2}4\right] } 
\end{equation}
where $E=E^{(i)}+\Delta _1^{(i)},\,\,\Gamma _1=\Gamma _{BW}\,\,$ and $%
\,\Gamma _2=\Gamma ^2/\Gamma _1$. The value of $\Gamma _2$ is found from the
exact relation (\ref{delE}). Note that $F_k^{(i)}$ from Eq. (\ref{newBW})
automatically satisfies another exact relation $\sum%
\nolimits_iE^{(i)} F_k^{(i)}=E_k$ ; note, that the contribution of the energy
shift $\Delta _1^{(i)}$ is compensated by the increase of the level density $%
\rho (E^{(i)})$. 

It follows from Eq.( \ref{newBW}) that the shift of energy due to the
enhanced admixture of higher basis components is two times smaller than in
Eq. (\ref{shift}) since the increase of density $\rho _0$ is partly
compensated by the factor $\rho _0{}^{-1/2}$ in Eq.(\ref{newBW}): 
\begin{equation}
\label{shiftnew}\Delta _2^{(i)} \simeq \frac{d(\ln \sqrt{%
\rho _0})}{dE}\overline{\left( \Delta E\right) ^2}\simeq \Delta _1^{(i)} 
\end{equation}
Thus we should substitute to Eq.(\ref{eqs}) the value $E=E^{(i)}+\Delta
_1^{(i)}+\Delta _2^{(i)}\simeq 2H_{ii}-E^{(i)}$. After taking into account
this shift, the FD expression gives the same result as the $F-$%
distribution; namely, in Fig.2b the FD curve (circles) is shifted
to that given by stars, both curves also agree with numerical experiment
(Fig.3). 

 For small $V_0\,$ and large number of particles we have $\Delta
_1^{(i)}\ll \Gamma _1\ll \Gamma _2$ , therefore, in the central part the
distribution (\ref{newBW}) has the Breit-Wigner shape with the width $\Gamma
_{BW}$ . However, numerical calculations \cite{ce}\cite{ZELE}\cite{zel2}
demonstrate that at larger interaction $V_0$ the width of the spreading
function becomes linear in $V_0$ and it is better to put $\Gamma _1=\Gamma
_2=\Gamma $. One can write the extrapolation expression both for small and
large values of $V_0$ (see also \cite{zel2}) $\Gamma _1=\Gamma _{BW}\Gamma
/(\Gamma _{BW}+\Gamma ).$ We have also checked that for large number $N_{pc}$
of principal components the distribution of the occupation numbers does not
depend on $\Gamma_1$; this fact can be treated as a signature of the
equilibrium. 

 Finally, we discuss the transition to mesoscopic systems. The result
depends on the dimensionality $d$ of the system since $d_0\sim
l^{-2};\,V_0\sim l^{-d}$ , therefore, $V_0/d_0\sim l^{-(d-2)}$ where $l$ is
the size of the system. Thus, for $d=1$ one has $V_0\gg d_0$ which means
that strong mixing (chaos) starts just from the ground state. This is in
accordance with the absence of a gap in the distribution of occupation
numbers in 1D case (Luttinger liquid). In the 3D case we have $V_0\ll d_0$
and the admixture of the higher states to the ground state can be considered
perturbatively which is consistent with the non-zero gap at $T=0$ . One can
see that the transition between regular region (I) and the initial chaos
region II in the 3D case occurs for high states when $V_0>D\sim d_0\exp
(-n_2^{*})$ where $n_2^{*}\propto E^{1/2}$ is a number of excited particles.
On the other hand, the transition from the fluctuating to the equilibrium
regime, (II)$\rightarrow $ (III), requires the decrease of $D$ only by one
order of magnitude: $n_3^{*}\approx n_2^{*}+2$ . This means that such
transition in 3D is a very sharp ``phase transition''. 

 In conclusion, we developed new method for the calculation of the
occupation numbers in finite systems of interacting particles. The method is
based on the assumption that exact eigenstates are ``chaotic''
superpositions of the shell-model basis states, and the smooth spreading
function for the eigenstates components can be introduced. This assumption
and the results are confirmed by numerical experiments. The
''microcanonical'' partition function which we have introduces can be used
for further studies of statistical and thermodynamic properties of finite
systems of interacting particles.

 We also demonstrated
that occupation numbers in the systems of interacting particles can be
reasonably described by Fermi-Dirac distribution with renormalized parameters.
As usual, mean field  (and possibly other ``regular'' effects) can be
included into single-particle energies $\tilde \epsilon_s$. Staistical effects
of the residual interaction (mainly due to non-diagonal matrix elements of the
Hamiltonian matrix) increase effective temperature.

 The authors are thankful to Y.Fyodorov, G.Gribakin, M.Kuchiev,
I.Ponomarev, V.Sokolov, O.Sushkov and V.Zelevinsky for the discussions;
 F.M.I is very grateful to the
staff of School of Physics, University of New South Wales for the
hospitality during his visit when this work was done. This work was
supported by the Australian Research Council.

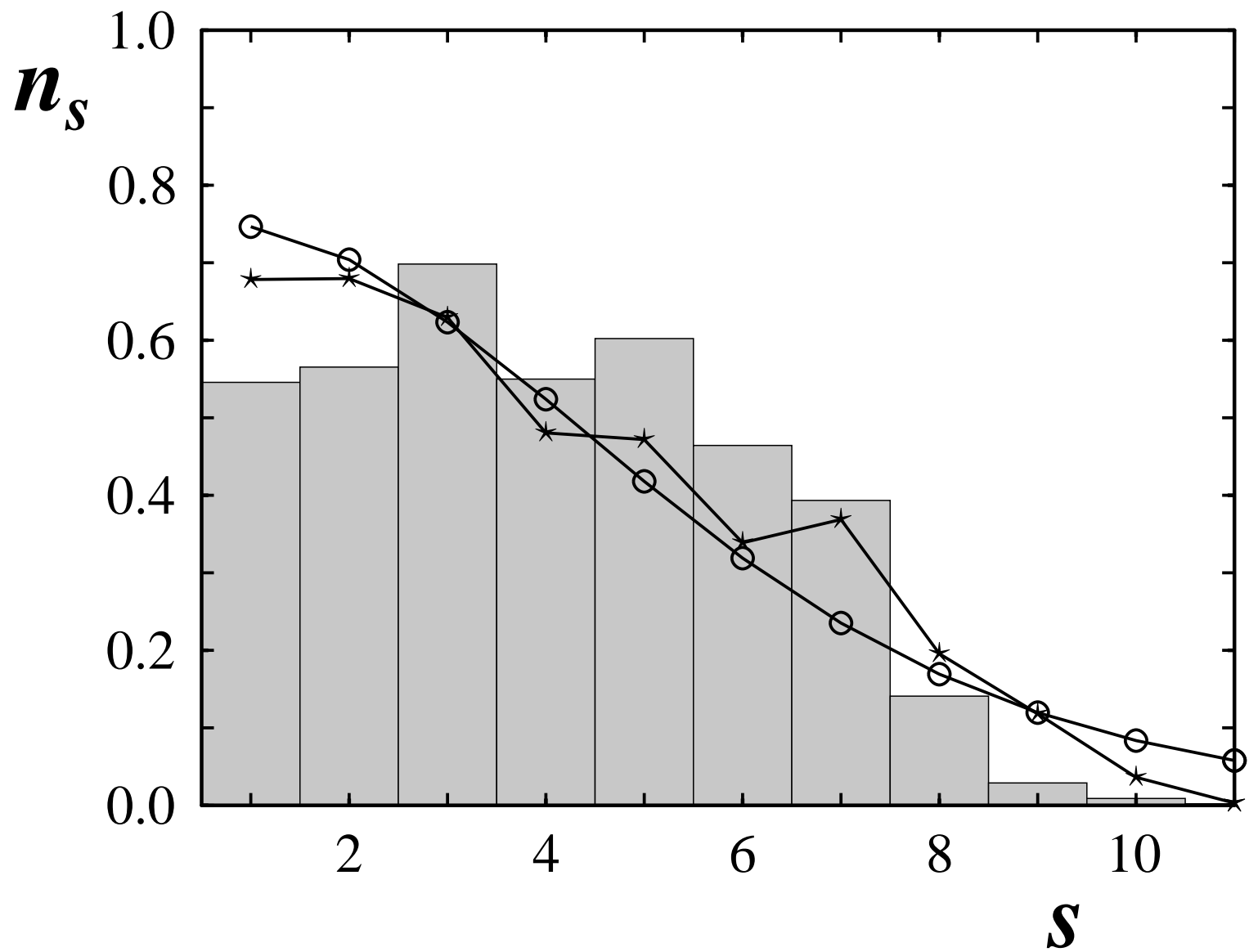
 Fig.1. Distribution of occupation numbers $n_s$ for $n=4$ particles
and $m=11$ orbitals for the two-body random interaction model \cite{FGI96}, 
\cite{FIC96} with $d_0=1$ , weak interaction $V_0=0.04$ and total energy $%
E=17.33$ (to compare with the ground state energy $E_F\approx 12.1)$ .
Dashed boxes represent numerical data averaged over $20\,$ Hamiltonian
matrices of the size $N=330$ with different realization of the random
interaction and over the energy range $E\pm \delta E$ with $\delta E=0.25$.
Circles correspond to the FD distribution with temperature $T$ and
chemical potential $\mu $ defined by total number $n$ of particles and total
energy $E$ . Stars are given by the $F-$ distribution (\ref{nalpha})
for $n_s(E)$ . The number of principal components in exact eigenstates is $%
N_{pc}\approx 13.0$ . The latter has been calculated via the inverse
participation ratio, $N_{pc}=(\sum\nolimits_kF_i^2(E_k-E))^{-1}$. Large
fluctuations $\sim N_{pc}^{-1/2}$ are seen. 

 Fig.2. Distribution of occupation numbers $n_s\,$for larger number $%
n=14$ of particles and $m=28$ of orbitals with the same $d_0$ and
temperature. Direct diagonalization of huge Hamiltonian matrices is not
possible in this case. Circles are the FD distribution; stars are
the $F-$distribution (\ref{nalpha}). 

 (a) weak interaction, $V_0=0.003,$ $\Gamma \approx 0.62$ , $%
N_{pc}\approx 60.$ 

 (b) strong interaction, $V_0=0.08,$ $\Gamma \approx 16.6$ , $%
N_{pc}\approx 11000.$ 

 Fig. 3 Distribution of occupation numbers for Ce atom parameters $%
n=4,m=11,d_0=1,V_0=0.12,N_{pc}\approx 48$ (boxes) in comparison with the
FD distribution, with the corrected temperature (circles) and with
the $F-$distribution (stars). 

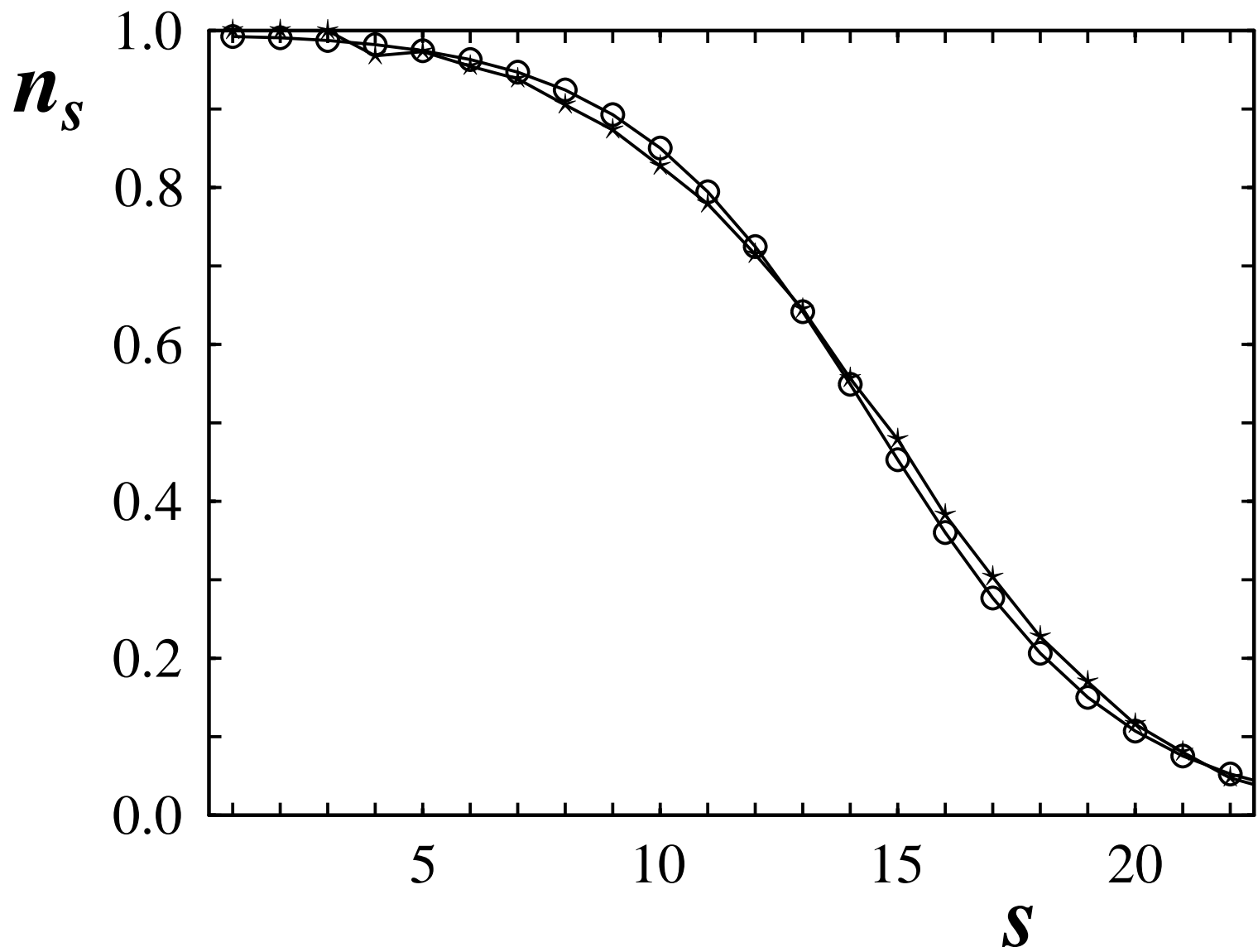
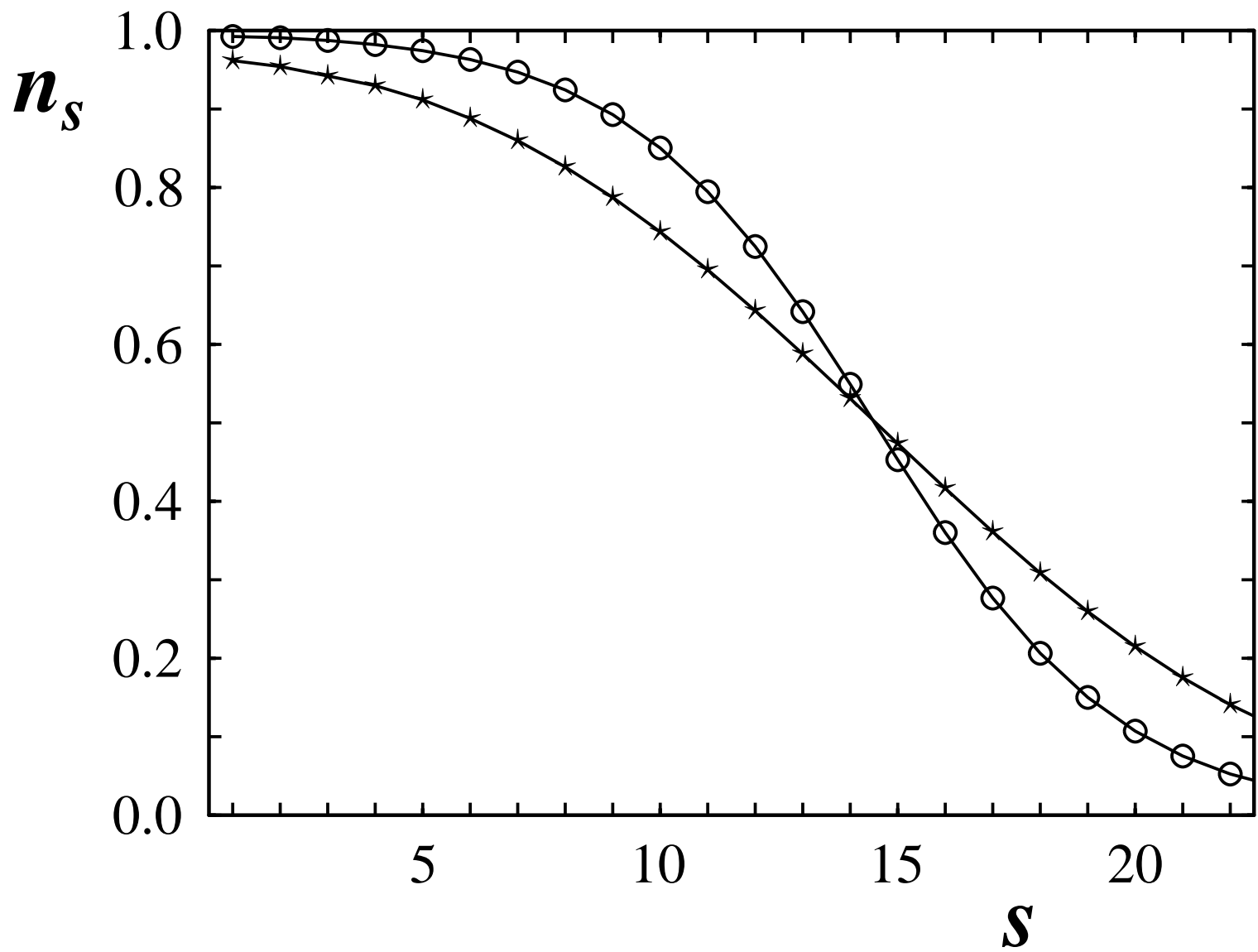
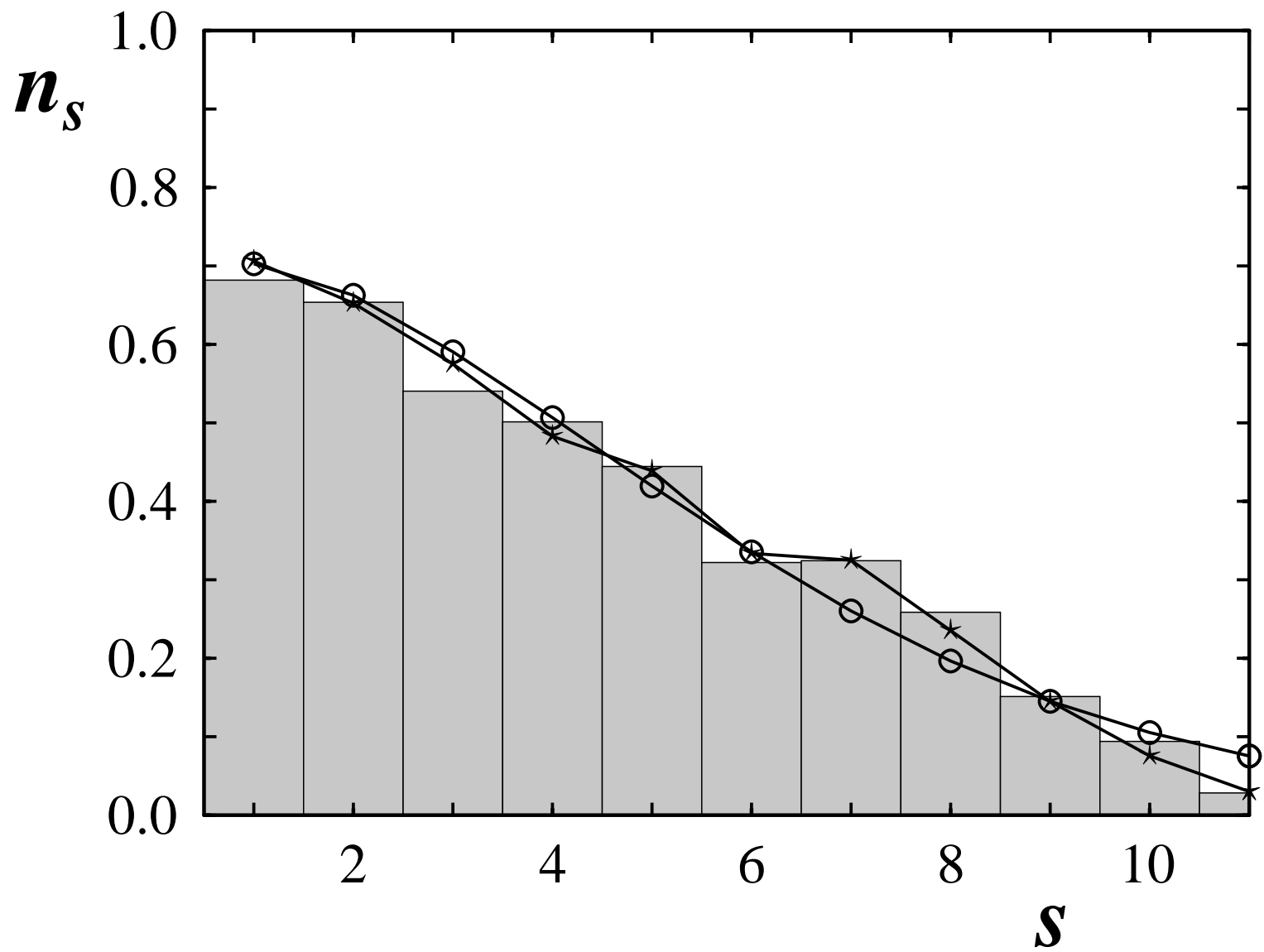
\end{document}